\documentclass{raa}            

\usepackage{graphicx,times}             

\begin{document}

   \title{Periodicity in fast radio bursts due to forced precession by a fallback disk
}

   \volnopage{Vol.0 (20xx) No.0, 000--000}      
   \setcounter{page}{1}          

   \author{H. Tong
      \inst{1}
   \and W. Wang
      \inst{2,3}
   \and H. G. Wang
      \inst{1}
   }

   \institute{School of Physics and Electronic Engineering, Guangzhou University,
   Guangzhou 510006, China; {\it htong\_2005@163.com}\\
\and School of Physics and Technology, Wuhan University, Wuhan 430072, China \\
\and WHU-NAOC Joint Center for Astronomy, Wuhan University, Wuhan 430072, China \\
   }

   \date{Received~~2009 month day; accepted~~2009~~month day}

\abstract{Recently, a 16-day periodicity in one fast radio burst is reported.
We propose that this 16-day periodicity may be due to forced precession of the neutron star
by a fallback disk. When the rotation axis is misaligned with the normal direction of the disk plane, the neutron star will precess. The eccentricity of the neutron star may be due to rotation or strong magnetic field etc. We found that the 16-day period may be understood using typical masses of the fallback disk. Polarization observations and information about the neutron star rotation period may help to discriminate different models. The possible precession observations in pulsars, magnetars, and fast radio bursts may be understood together using forced precession by a fallback disk.
}

   \authorrunning{Tong, Wang, \& Wang}            
   \titlerunning{Periodicity in FRBs due to forced precession}  

   \maketitle

%
%
\section{Introduction}           

Recently, a periodicity of 16 days in one fast radio burst is reported (FRB 180916.J0158+65, Amiri et al. 2020). If the observation is reliable, it opens a new channel to explore the nature of fast radio bursts (Lorimer et al. 2007; Thornton et al. 2013). This 16-day periodic signature may correspond to a binary orbital period (Lyutikov et al. 2020; Yang \& Zou 2020 (forced precession by a binary companion); Ioka \& Zhang 2020). In some of the models, repeated fast radio bursts are thought to originate from activities of isolated magnetars. In the isolated magnetar case, the 16-day periodicity may be caused by free precession of the magnetar (Levin et al. 2020; Zanazzi \& Lai 2020). We propose that the 16-day periodicity may be due to forced precession of an isolated neutron star by a fallback disk.

In previous studies of pulsars and magnetars, the forced precession by a fallback disk has already been discussed (Qiao et al. 2003 and references therein). Free precession in normal neutron stars is thought to be impossible due to the liquid core of neutron star (Shaham 1977). When the possible precession signal was seen in PSR B1828$-$11 (Stairs et al. 2000), it was argued about whether it is due to free precession (Link \& Epstein 2001) or forced precession by a fallback disk (Qiao et al. 2003).

Magnetars may be young neutron stars with very high magnetic field (Duncan \& Thompson 1992). One alternative to the magnetar model is the fallback disk model (Katz et al. 1994; Chatterjee et al. 2000; Alpar 2001). And one fallback disk is indeed observed in the magnetar 4U 0142+61 (Wang et al. 2006). Furthermore, in 4U 0142+61 a possible signal of precession was also reported (Makishima et al. 2014, 2019), which may be due to free precession of the central magnetar (Makishima et al. 2019). Considering that the magnetar has a fallback disk, it is possible that the precession may also occur due to forced precession by the fallback disk (like that in PSR 1828$-$11). A magnetar with a very long rotation period of 6.6 hours was reported (D'Ai et al. 2016; Rea et al. 2016). And this long rotation period could be also due to the effect of a fallback disk (Tong et al. 2016).

Therefore, from previous experiences in pulsars and magnetars, the presence of a fallback disk can explain long periodicity observations. This can be achieved by either forced precession or interaction between the central neutron star and the fallback disk. The pulsar PSR B1828$-$11 has a pulsation period of $0.4$ seconds and a possible precession period about $1000$ days (Stairs et al. 2000; Ashton et al. 2017). If the central neutron star inside fast radio bursts has a smaller period, then it is generally expected that the corresponding precession period is also shorter. This is qualitatively consistent with the observations of a 16-day period in FRB 180916.J0158+65 (Amiri et al. 2020). Quantitative calculations and discussions are presented below.

\section{Calculation of the period of forced precession}

\subsection{Eccentricity due to rotation}

The detailed modeling of neutron star forced precession by a fallback disk can be found in Qiao \& Cheng (1989), Qiao et al. (2003).
The geometry of the neutron star and the fallback disk is shown in figure 1 in Qiao et al. (2003). In the geometry depicted in Qiao et al. (2003), the rotational axis is aligned with one of the principle axes of moment of inertial. Therefore, there is no free precession. Only forced precession by the disk is possible. The main result is: the angular velocity of the forced precession is (equtaion (2) in Qiao et al. 2003, here only the absolute value is needed)
\begin{equation}
  |\dot{\phi}| = \frac{3 G M_0 \cos\theta}{2 c d (d+c) \Omega} (1-\frac{b^2}{a^2}),
\end{equation}
where $G$ is the gravitational constant, $M_0$ is the total mass of the fallback disk, $\theta$ is the angle between the neutron star rotation axis and the normal direction of the disk plane, $c$ and $d$ are the inner/outer radii of the disk, respectively, $\Omega$ is the rotational angular velocity of the neutron star (denoted as $\omega$ in Qiao et al. 2003), $a$ and $b$ are the radii of the semi-major and semi-minor axis of the neutron star (which is deformed by rotation or magnetic field etc). Denote $M_{\theta}\equiv M_0 \cos\theta$, since $M_0$ and $\cos\theta$ always appear together. The inner and outer disk radii may be of the same order, denoted as  $c \sim d \sim R$. The eccentricity of the neutron star is defined as $e =(1-b^2/a^2)^{1/2}$ (Shapiro \& Teukolsky 1983). Then the angular velocity of the forced precession is
\begin{equation}
  |\dot{\phi}| = \frac{3 G M_{\theta}}{4 R^3 \Omega} e^2.
\end{equation}
Similar expression is also obtained when considering forced precession for PSR B1828$-$11 by a planet (Liu et al. 2007).

\begin{figure}
\centering
\includegraphics[scale=0.9]{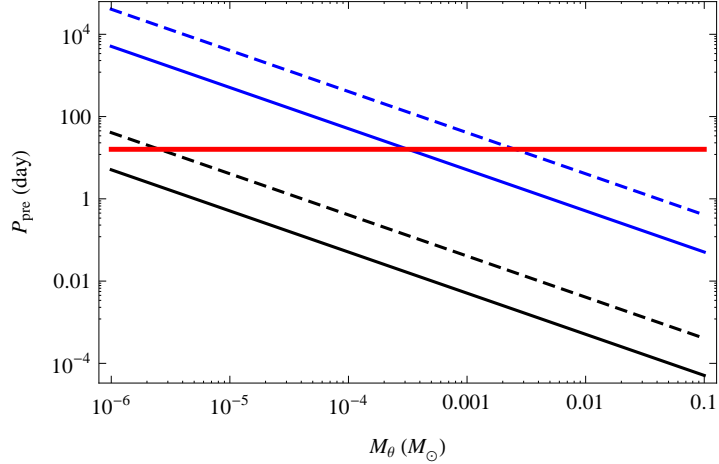}
\caption{\label{gmodel1}
Precession period as a function of the disk mass. The black solid line is for a neutron star rotation period of $5$ ms, the black dashed line is for a neutron rotation period of $10$ ms. For both of the black lines, the distance from the neutron star to the disk is taken as the corotation radius. The blue lines are similar to the black lines, except that the distance is taken as $10$ times the corotation radius. The red horizontal line is the observation of FRB 180916.J0158+65 (Amiri et al. 2020).}
\end{figure}

If the eccentricity of the neutron star is due to rotation, using the Maclaurin spheroids approximation, the eccentricity is related to the rotational angular velocity of the neutron star (Shapior \& Teukolsky 1983; Zhou et al. 2014)
\begin{equation}
  \Omega = 2e \sqrt{\frac{2\pi \rho G}{15}},
\end{equation}
where $\rho$ is the mean density of the neutron star.
From neutron star accretion disk modeling, the typical distance between the neutron star and the disk should be order of the corotation radius (Shapiro \& Teukolsky 1983; Qiao et al. 2003; Tong \& Wang 2019)
\begin{equation}
  R = \kappa R_{\rm co}
\end{equation}
where $\kappa$ is dimensionless parameter, $R_{\rm co} = (GM_{\rm ns}/\Omega^2)^{1/3}$ is the corotation radius, $M_{\rm ns}$ is the mass of the neutron star. Then the period of the forced precession by a fallback disk is
\begin{eqnarray}
  P_{\rm pre} &=& \frac{2\pi}{|\dot{\phi}|} \\
  &=& \frac{2 G M_{\rm ns}}{15 \pi^2 R_{\rm ns}^3} \frac{M_{\rm ns}}{M_{\theta}} \kappa^3 P_{\rm ns}^3\\
  &=& 0.5 \kappa^3 \left( \frac{M_{\theta}}{10^{-5} M_{\odot}} \right)^{-1}
  \left( \frac{P_{\rm ns}}{5 \ \rm ms} \right)^3 \ {\rm day},
\end{eqnarray}
where $R_{\rm ns}$ is the radius of the neutron star, and $P_{\rm ns}$ is the rotational period of the neutron star. Typical values of the precession period is given for typical values of fallback disk mass and rotational period of the neutron star. The mass of the fallback disk may range from $(10^{-6}-0.1) \ \rm M_{\odot}$ (Wang et al. 2006; Perna et al. 2014). The rotational period of a newly born magnetar may in the order of $5\ \rm ms$ (Vink \& Kuiper 2006; Zhou et al. 2019). For several combinations of neutron period and disk radii, the precession period as a function of the disk mass is shown in figure \ref{gmodel1}.
It can be seen that the 16-day period in FRB 180916.J0158+65 can be understood naturally for typical masses of the fallback disk.

\subsection{Eccentricity due to other origins}

\begin{figure}
\centering
\includegraphics[scale=0.9]{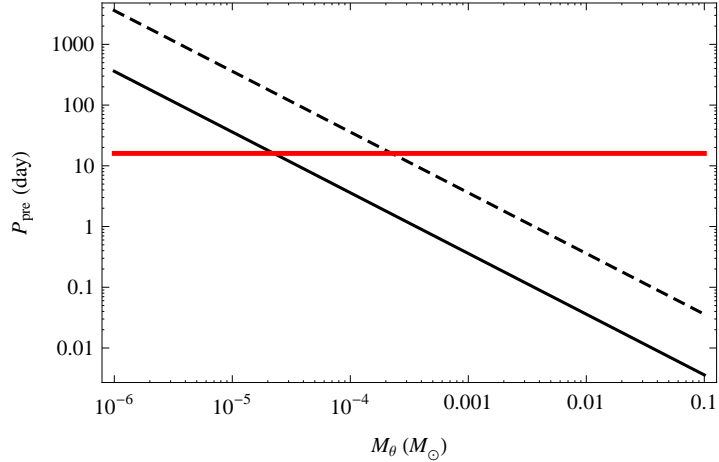}
\caption{\label{gmodel2}
Precession period as a function of the disk mass, when the eccentricity is due to other origins. The black solid line is for an ellipticity of $10^{-4}$, the black dashed line is for an ellipticity of $10^{-5}$.  For both of the black lines, the distance from the neutron star to the disk is taken as the corotation radius and the neutron star rotation period is taken as $5$ ms. The red horizontal line is the observation of FRB 180916.J0158+65 (Amiri et al. 2020). }
\end{figure}

In Qiao et al. (2003), they only considered the deformation of the neutron star by the effect of rotation.
In principle, other factors can also deform the neutron star (Zanazzi \& Lai 2020 and references therein). Magnetars may have strong magentic field about $(10^{14} -10^{15}) \ \rm G$ (Kaspi \& Beloborodov 2017). Their internal toroidal magnetic field may be even stronger, e.g. may be as high as $10^{16} \ \rm G$. This strong toroidal magnetic field will also deform the neutron star (Usov 1992; Makishima et al. 2019). The ellipticity (fractional difference of moment of inertial along different axes) of the neutron star is about (Makishima et al. 2019)
\begin{equation}
  \varepsilon \sim 10^{-4} \left( \frac{B_{\rm t}}{10^{16} \ \rm G} \right)^2,
\end{equation}
where $B_{\rm t}$ is the internal toroidal magnetic field.
The relation between the eccentricity and ellipticity is: $e^2 = 3\varepsilon$ (Shapiro \& Teukolsky 1983).
Therefore, if the deformation of the neutron star is due to other origins, the period of the forced precession by a fallback disk is
\begin{eqnarray}
  P_{\rm pre} &=& \frac{4}{9\varepsilon} \frac{M_{\rm ns}}{M_{\theta}} \kappa^3 P_{\rm ns}\\\label{eqn_Ppre_model2}
  &=& 36 \kappa^3 \left( \frac{\varepsilon}{10^{-4}} \right)^{-1}
  \left( \frac{M_{\theta}}{10^{-5} \ \rm M_{\odot}} \right)^{-1}
  \left( \frac{P_{\rm ns}}{5 \ \rm ms} \right) \ {\rm day}.
\end{eqnarray}
The corresponding precession period as a function of disk mass is shown in figure \ref{gmodel2}. Again, the 16-day period in FRB 180916.J0158+65 can be understood using typical values of the fallback disk mass.

\section{Discussion}

\subsection{How to discriminate between different models?}

The 16-day period in FRB 180916.J0158+65 may be due to orbital origin or the neutron star itself. In the case of pulsars, change of polarization is the direct evidence of a rotating neutron star and a rotating magnetic field (Rahhakrishnan \& Cooke 1969; Lyne \& Manchester 1988). Therefore, if future polarization information of the periodically active fast radio burst can also be obtained, the orbital origin and neutron star precession origin  may be distinguished.

If only a small region on the neutron star surface is responsible for the radio emission, then a precessing neutron star may explain the phase dependence (4-day phase window) of the radio burst (Zanazzi \& Lai 2020). This is mainly a geometrical effect. It does not dependent on whether the precession is free or forced. If the 16-day period is due to precession of the neutron star, it can be either free precession or forced precession. For free precession, it may be damped by superfluid and viscosity of the liquid core (Shaham 1977; Sedrakian et al. 1999). For the possible precession in PSR B1818$-$11 (Stairs et al. 2000), Qiao et al. (2003) considered forced precession as an alternative. Similarly, this is also the reason that we propose a forced precession model for the 16-day periodicity in FRB 180916.J0158+65.

For free precession, the precession period is about $P_{\rm ns}/\varepsilon$ (Makishima et al. 2019). For an ellipticity of $\varepsilon \sim 10^{-4}$, in order to have a 16-day precession period, the required neutron star rotation period should be of the order of $\sim 100 \ \rm s$. Or a small ellipticity is required even in the presence of strong internal magnetic field (Levin et al. 2020). It is not sure whether this can be accomplished in theory. While, in the forced precession scenario, we use typical values of $P_{\rm ns} \sim 5 \ \rm ms$ for new born neutron stars (which can be either a normal pulsar or a magnetar). If in the future some information about the the rotation period of the central neutron star can be obtained, then the free or forced precession models may be constrained by the observations. If the corresponding neutron star period is about 1 second, then the the disk forced precession is also valid. The typical disk mass may change, see equation (\ref{eqn_Ppre_model2}). At present, the rotation period of the neutron star is a free parameter.

Here, the fast radio burst is assumed to originate from a neutron star. The young neutron star can be formed by core-collapse of massive stars, binary neutron star mergers etc. In different formation channels, the disk properties are distinct. Moreover, if the young neutron star is formed by core-collapse (Zhang et al. 2020), it should be associated with supernova explosion and/or gamma-ray bursts (Zhang 2014). The corresponding fallback disk properties may be similar to that in the Galaxy. For isolated neutron stars, fast radio burst may occur in the magnetosphere (Yang \& Zhang 2018; Lyutikov 2019) or far away from the neutron star (Metzger et al. 2019; Beloborodov 2019). Around the neutron star, there are some fallback material in the form of a disk. If the fast radio burst is a neutron star with some asteroids or other small planets (Dai et al. 2016), the dynamics may be similar to the fallback disk case (Liu et al. 2007). If the neutron star is a rigid body as a whole (e.g. a solid quark star, Xu 2003), the discussion about free precession and origin of deformation will be totally different. There are also internal origins that can generate perturbations about tens of days (e.g., Tkachenko modes, Noronha \& Sedrakin 2008). But how the internal perturbations can affect the radio appearance and disappearance is unknown.

\subsection{Neutron star+fallback disk systems}

The period of possible precession in PSR B1828$-$11 is about $1000$ days (Stairs et al. 2000; Ashton et al. 2017). In the magnetar with fallback disk 4U 0142+61, the precession period is about $0.5$ days (Makishima et al. 2019). Whether and how fallback disks are formed is still an open question (Perna et al. 2014). The magnetar 4U 0142+61 is found to have a possible fallback disk (Wang et al. 2006). The characteristic age of 4U 0142+61 is about $6.8\times 10^{4} \ \rm yr$ (Olausen \& Kaspi 2014). In the literatures, the typical life of the fallback disk is about $2\times 10^{4} \ \rm yr$ (Menou et al. 2001; Li 2007). The possible fallback disk in fast radio bursts can also operate for this typical duration. In the fast radio burst case, the period is about $16$ days (Amiri et al. 2020). From the above modeling, the precession period depends on the neutron star rotation period, i.e. whether it is a new born neutron star or slowed-down neutron star. The precession period also depends sensitively on the typical radii of the fallback disk, whether it is order of or much higher than the corotation radius. It also depends on the mass and geometry of the fallback disk $M_{\theta}$. Therefore, there is a large parameter space for the forced precession of neutron stars by a fallback disk. The 16-day period in FRB 180916.J0158+65 can be understood using typical parameters of the fallback disk. Later observations of more quasi-periodic phenomena in pulsars (Kramer et al. 2006; Lyne et al. 2010) may favor a magnetospheric origin instead of a precession origin. However, why the magnetosphere is moulded quasi-periodically is unknown. And forced precession due to an external fallback disk is still one candidate for moulding the magnetosphere quasi-periodically.

Here, we only consider the dynamical effect of the fallback disk. The interaction between the fallback disk and the neutron star (accretion or propeller) will make the final output more diverse. In the case of pulsars and magnetars, the presence of a fallback disk may (1) cause the pulsar to null for part of the time (Li 2006), (2) explain the braking index of pulsars (Liu et al. 2014), (3) force the neutron star to precess (as discussed above), (4) provide an alternative model to magnetars to explain various pulsar-like objects (Alpar 2001), (5) explain the rotational evolution of the magnetar with a rotation period of 6.6 hours (Tong et al. 2016), etc. Therefore, the problem of fallback disks (including forced precession due to the fallback disk) in pulsars, magnetars, and fast radio bursts needs more studies in the future.

\textit{Notes added: After this paper is submitted, a possible 159-day period is reported in the repeating FRB 121102 (Rajwade et al. 2020, arXiv:2003.03596). This 159-day period can also be explained using the fallback disk model, see figure \ref{gmodel1} and \ref{gmodel2}.}

\section*{Acknowledgments}
The authors would like to thank the referee very much for helpful suggestions. H.Tong is supported by NSFC (11773008). W. Wang is supported the National Program on Key Research and Development Project (Grants No. 2016YFA0400803) and the NSFC (11622326 and U1838103). H.G. Wang is supported by NSFC (11573008) and the 2018 Project of Xinjiang Uygur Autonomous Region of China for Flexibly Fetching in Upscale Talents.

\label{lastpage}

\end{document}